\documentclass[conference]{IEEEtran}
\IEEEoverridecommandlockouts
\usepackage{cite}
\usepackage{amsmath,amssymb,amsfonts}
\usepackage{algorithmic}
\usepackage{graphicx}
\usepackage{textcomp}
\usepackage{xcolor}
\usepackage{multirow}
\usepackage{array}
\usepackage{multicol}
\usepackage{etoolbox}
\usepackage{bm}
\usepackage{subfig}
\usepackage{etoolbox}
\makeatletter
\patchcmd{\@makecaption}
  {\scshape}
  {}
  {}
  {}
\makeatother
\DeclareMathOperator*{\argmax}{arg\,max}
\def\BibTeX{{\rm B\kern-.05em{\sc i\kern-.025em b}\kern-.08em
    T\kern-.1667em\lower.7ex\hbox{E}\kern-.125emX}}

\begin{document}

\title{Identifying Protein-Protein Interaction using Tree LSTM and Structured Attention\\
{\footnotesize \textsuperscript{}}
}
    \author{\IEEEauthorblockN{Mahtab Ahmed, Jumayel Islam, Muhammad Rifayat Samee, Robert E. Mercer}
\IEEEauthorblockA{\textit{Department of Computer Science} \\
\textit{University of Western Ontario}\\
London, Ontario, Canada \\
mahme255, jislam3, msamee, rmercer@uwo.ca}}

\maketitle

\begin{abstract}
Identifying interactions between proteins is important to understand underlying biological processes. Extracting a protein-protein interaction (PPI) from the raw text is often very difficult. Previous supervised learning methods have used handcrafted features on human-annotated data sets. In this paper, we propose a novel tree recurrent neural network with structured attention architecture for doing PPI. Our architecture achieves state of the art results (precision, recall, and F1-score) on the AIMed and BioInfer benchmark data sets. Moreover, our models achieve a significant improvement over previous best models without any explicit feature extraction. Our experimental results show that traditional recurrent networks have inferior performance compared to tree recurrent networks for the supervised PPI problem.

\end{abstract}

\begin{IEEEkeywords}
Protein-Protein Interaction, Bioinformatics, Tree LSTM, Structured Attention
\end{IEEEkeywords}

\section{Introduction}

With extensive ongoing research currently happening in the bio-medical field, there is an exponentially growing amount of information available in textual form requiring expert knowledge to extract the important information contained therein. As doing this manually with human expertise only is time consuming and expensive, there has been a lot of interest in developing computational approaches for automatically inferring some hidden information from this vast source of knowledge such as protein-protein interactions (PPIs), drug-drug interactions (DDIs) and chemical-disease relation information. Researchers have successfully applied natural language processing (NLP) techniques and machine learning (ML) methods for doing these tasks \cite{peng2016improving, singhal2016text, huang2015community, leaman2016taggerone}.

The task of identifying protein-protein interactions (PPIs) is to extract relations between protein entities mentioned in a document  \cite{krallinger2008overview}. While PPI relations can cross over sentences and even across corpora, current work is centered mostly on PPIs in single sentences \cite{pyysalo2008comparative, tikk2010comprehensive}. For example, in the sentence ``\textit{LEC induces chemotaxis and adhesion by interacting with CCR1 and CCR8.}'', \textit{LEC}--\textit{CCR1} and \textit{LEC}--\textit{CCR8} are in PPI relations, whereas there is no relation between \textit{CCR1} and \textit{CCR8}.

Whereas previously, pattern-based methods have been very popular for doing this bio-medical relation extraction, in this paper, we propose a novel neural net architecture for identifying protein-protein interactions from bio-medical text using a Tree LSTM \cite{tai2015improved} with Structured Attention \cite{bahdanau2014neural}. We provide an in depth analysis of traversing the dependency tree of a sentence through a child sum tree LSTM and at the same time learn this structural information through a parent selection mechanism by modeling non-projective dependency trees. We also provide an extensive evaluation of our model by doing a detailed comparison with the currently available state of the art methods applied on the standard PPI corpora (AIMed, BioInfer, IEPA, HPRD50, and LLL). Our architecture achieves state of the art results on four of the five corpora. Our experiments suggest that our model is more generalized and is better capable of capturing long distance information than existing feature and kernel based methods.

\section{Related Work}
In previous work, pattern-based methods have been very popular for doing PPI relation extraction, where patterns as well as rules were crafted and defined based on lexical and syntactic features \cite{corney2004biorat , segura2011linguistic, leeuwenberg2015exploring}. For example, Leeuwenberg et al. \cite{leeuwenberg2015exploring} propose the syntactic tree pattern structure (STPS) for DDI extraction from a sentence in bio-medical text based on the syntax tree of the sentence. Also much research has been done on bio-medical relation extraction using Kernel-based methods which allow learning rich structural data in the form of syntactic parse trees and dependency structures \cite{miwa2009rich, kim2010walk, chang2016pipe, airola2008all}. Miwa et al. \cite{miwa2009rich} propose a system which embeds rich feature vectors in a Support Vector Machine with corpus weighting where the weights are learned from one corpus and the other corpora are used for support. Kim et al. \cite{kim2010walk} propose a walk-weighted sub-sequence kernel for the extraction of PPIs. It captures the non-contiguous syntactic structures by matching the v-walk and e-walk on the shortest dependency path. Chang et al. \cite{chang2016pipe} propose an interaction pattern tree kernel method in which they extract PPIs by integrating the PPI patterns with a convolution tree kernel. Airola et al. \cite{airola2008all} propose a method to extract PPIs by looking at the information from both dependency as well as linear subgraphs. For this they adopted an all-path kernel approach where they weighted all the edges on the shortest paths by a high value and all other edges with a low value. Peng et al. \cite{peng2015extended} propose an Extended Dependency Graph (EDG) based approach by incorporating a few simple linguistic features beyond syntax information. Finally they evaluated this EDG approach with edit distance and an APG kernel on the five benchmark corpora. Zhang et al.
\cite{zhang2011neighborhood} propose a neighborhood hash kernel based method for PPI extraction. They started by transforming each node label of the dependency graph for two target sentences into a bit label and then replaced this bit label by a new label produced by order-independent logical operations on the bit labels of the current node and its neighboring nodes. They continued this process for the two target sentences and finally ended up with a high order substructures over the dependency graph. Finally, they computed the similarity of the two dependency graphs based on the intersection ratio of the updated label sets.

Recently, deep neural network (DNN) based methods have successfully applied and achieved promising results in bio-medical relation extraction from bio-medical literature \cite{quan2016multichannel, liu2016drug,zeng2014relation, zhao2016protein}. Mikolov et al. \cite{mikolov2013distributed} propose an approach which gives a distributed representation (i.e., embeddings) of words capturing both the syntactic and the semantic similarity. Nowadays almost all of the DNN-based approaches in linguistics have this embedding layer at the top either in a pre-trained or randomly initialized form.

Peng et al. \cite{peng2017deep} adopt a convolutional neural network (CNN) based approach in which they utilize two channels of CNN for high level feature extraction. In one channel, they use raw words along with some syntactic features such as parts-of speech, chunk parsing information, named entities, syntactic dependencies and two distance vectors for each word representing the distance from the word to the two proteins being considered as interacting. In another channel they use parent word information for each word and pass this to an embedding layer to get a distributed representation of the sentence in terms of parent words. Following this, they apply convolution on these two channels separately and map them via a fully connected layer to the required number of classes.

Zhang et al. \cite{zhang2018hybrid} also utilize a multi-channel CNN for this task. In the first channel, they use a sequence of raw words along with positional embedding features. In the second channel, they use shortest dependency path information: the arcs visited in the shortest paths from the first protein to the root and the second protein to the root. This information is arranged as a sequence and passed to an embedding layer. In the third channel, they use dependency relation embedding, storing the embedding of the words encountered in the shortest dependency path. Finally they apply convolution on it and map the extracted features to two classes using multi-layer perception (MLP) followed by a softmax layer.  

Zhao et al. \cite{zhao2016protein} propose a greedy layer-wise unsupervised learning-based approach to extract PPIs from bio-medical literature. They first divided their corpus into train, validation and unlabelled set and applied an auto encoder (AE) on the unlabelled set of data to initialize the parameters of a deep multi-layer neural network. Finally they applied a gradient descent method using back propagation to train their whole model.

Hsieh et al. \cite{hsieh2017identifying} utilize recurrent neural network (RNN) for extracting the PPI form bio-medical literature. Without doing any additional feature extraction, they used long short term memory (LSTM) (a variant of RNN) to encode the dependency information through time from forward and backward direction over the sentence. Finally, they took the left most and right most output vector of LSTM, concatenated them and applied MLP followed by softmax for the classification.

\begin{figure*}%
    \centering
    \subfloat[]{{\includegraphics[width=8cm,height=5 cm]{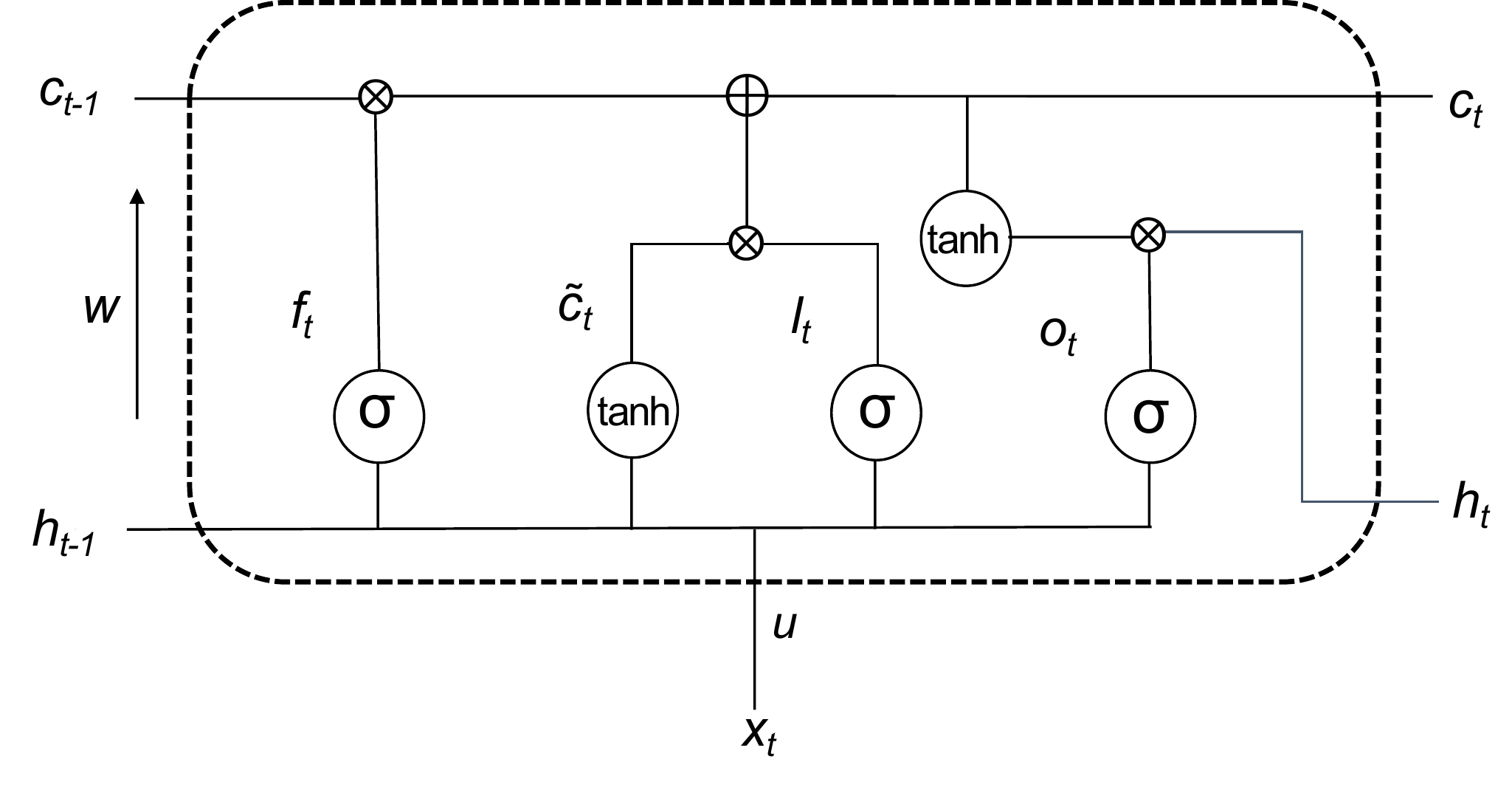} }}%
    \qquad
    \subfloat[]{{\includegraphics[width=9cm,height=5 cm]{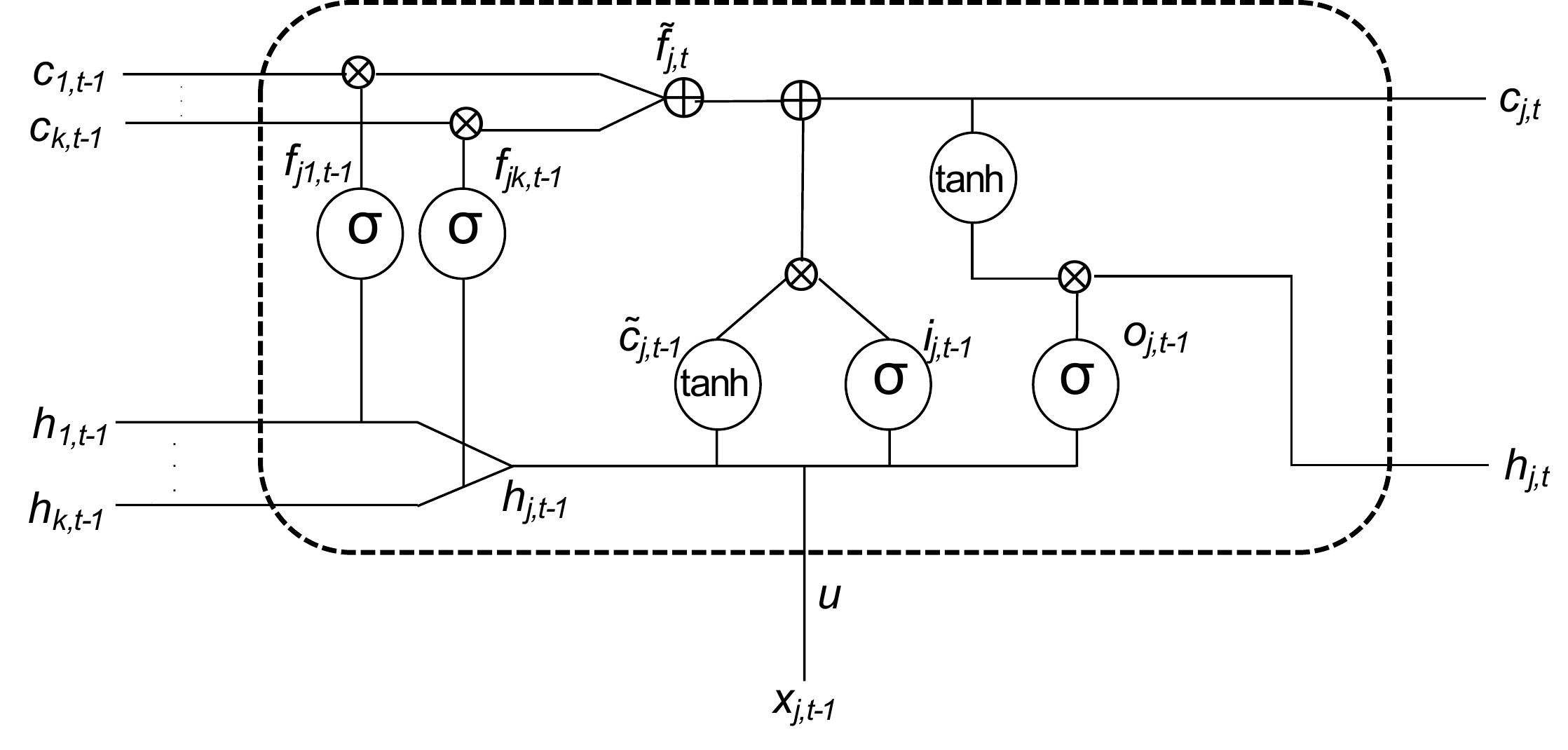} }}%
    \caption{Standard LSTM vs Tree LSTM}%
    \label{lstm}%
\end{figure*}

\section{The Model}

In this section, we describe our work in detail. We first explain the working mechanism of a tree LSTM cell. Then we explain the Structured Attention mechanism for learning the dependency tree through Kirchhoff’s Matrix-Tree Theorem. Finally we explain how we combine the tree LSTM architecture with structured attention to obtain a performance boost that we describe in the next section.

\subsection{ Recurrent unit: Bidirectional LSTM}
In this paper, we use recurrent neural network (RNN) which is the best known and most widely used NN model for sequence data. Its long-short term memory variant (LSTM) gives all of the advantages of the basic RNN with an elegant solution to RNN's vanishing gradient problem. Fig \ref{lstm}(a) shows a sample LSTM cell and the construction of its internal gates are as follows:
$$ \textbf{i}_t = \sigma (\textbf{W}^{(i)} x_t + \textbf{U}^{(i)} h_{t-1} + \textbf{b}^{(i)})$$
$$ \textbf{o}_t = \sigma (\textbf{W}^{(o)} x_t + \textbf{U}^{(o)} h_{t-1} + \textbf{b}^{(o)})$$
$$ \textbf{f}_t = \sigma (\textbf{W}^{(f)} x_t + \textbf{U}^{(f)} h_{t-1} + \textbf{b}^{(f)})$$
$$ \Tilde{\textbf{c}}_t = \textit{tanh} (\textbf{W}^{(c)} x_t + \textbf{U}^{(c)} h_{t-1} + \textbf{b}^{(c)})$$
$$ \textbf{c}_t = i_t \cdot \Tilde{c}_t + f_t \cdot c_{t-1} $$
$$ \textbf{h}_t = o_t \cdot {tanh(c_t)}$$

Although LSTMs are very good with sequence data, most often it is important to have information from the past as well as from the future. However, LSTM allows only one hidden state from the past and changes that hidden state recursively through time. An elegant resolution to this problem is going over the sequence in both forward and backward directions using two hidden states and finally concatenating the output from both directions. This method, called Bidirectional LSTM (BLSTM), has proven to be very effective in some prior works \cite{graves2005framewise, thireou2007bidirectional,  dyer2015transition}. BLSTM has the same internal structure as LSTM except one of the output dimensions is twice that of the LSTM output.

\subsection{Tree LSTM}
The main limitation of the basic LSTM is that it can only be used for analyzing sequential information. However, a natural language sentence encodes more than a sequence of words. This extra information is usually represented in a tree structure. One such structure is the dependency tree \cite{chen2014fast}. LSTM and BLSTM cannot analyze this structured information correctly. A variant of standard LSTM cell, called tree LSTM (tLSTM) \cite{tai2015improved}, traverses the sentence by following a tree-structured network topology rather than going over the sequence as a linear chain. The underlying idea of an LSTM cell remains the same except here each tLSTM unit is capable of incorporating information from multiple child units as well. Fig \ref{lstm}(b) shows a sample tLSTM cell. In this study, we use child sum version of tree LSTM, as it is more suitable with dependency trees.

Traditional LSTM takes the previous hidden state $h_{t-1}$, the previous cell state $c_{t-1}$ and the current time step input $x_t$ into account and generates a new hidden state and cell state. However in the child sum tree LSTM, the main gist remains the same except component node states are now generated based on the states of its all possible children in the tree structure. To do this, first, the hidden states at the previous time step is summed up for all of the children of the component node and the internal gates (i.e., input, output and intermediate cell state) are updated using this new hidden state. 
\begin{equation}
    \Tilde{\textbf{h}}_{j,t} = \sum_{k \in C(j)}^{}h_{jk, t-1}
\end{equation}
where $C(j)$ denotes the set of children of node $j$. Next using this modified hidden state $\Tilde{h}$, input, output and intermediate cell states are calculated as follows,
\begin{equation}
    \textbf{i}_{j,t} = \sigma (\textbf{W}^{(i)} x_{j,t} + \textbf{U}^{(i)} \Tilde{h}_{j,t} + \textbf{b}^{(i)})
\end{equation}
\begin{equation}
    \textbf{o}_{j,t} = \sigma (\textbf{W}^{(o)} x_{j,t} + \textbf{U}^{(o)} \Tilde{h}_{j,t} + \textbf{b}^{(o)})
\end{equation}
\begin{equation}
    \Tilde{\textbf{c}}_{j,t} = \textit{tanh} (\textbf{W}^{(c)} x_{j,t} + \textbf{U}^{(c)} \Tilde{h}_{j,t} + \textbf{b}^{(c)})
\end{equation}
where $W^{(i)}$, $W^{(o)}$ and $W^{(c)}$ are the parameters to be learned. Instead of having just a single forget gate, tLSTMs have $k$ forget gates where $k$ is equal to the number of children of the target node. This multiple forget gate allows tLSTM to incorporate individual information from each of the children in a selective manner. Each forget gate is calculated as follows: 
\begin{equation}
    \textbf{f}_{jk,t} = \sigma (\textbf{W}^{(f)} x_{j,t} + \textbf{U}^{(f)} h_{jk,t-1} + \textbf{b}^{(f)})
\end{equation}
Next, the individual forget gate outputs are multiplied with corresponding cell state values and then combined to get a single forget vector which is further used to get the final cell state of the model as follows: 
\begin{equation}
    \tilde{\textbf{f}}_{j,t} = \sum_{k \in C(j)}^{}f_{jk,t} \cdot {c_{k,t-1}}
\end{equation}
\begin{equation}
    \textbf{c}_{j,t} = i_{j,t} \cdot \tilde{c}_{j,t} + \tilde{f}_{j,t}
\end{equation}
Finally, the update equation for the hidden state of a child sum tree LSTM cell is similar to the one used in traditional LSTM,
\begin{equation}
    \textbf{h}_{j,t} = o_{j,t} \cdot \textit{tanh}(c_{j,t})
\end{equation}

Each of the parameter matrices represents a correlation among the component vector, input $x_j$ and the hidden state $h_k$ of the $k^{th}$ child of the component unit. For example, the \texttt{sigmoid} function at the input gate represents semantically important words at input by giving values close to 1 (e.g., a verb) and relatively unimportant words by giving values close to 0 (e.g., a determiner). Since the hidden state and cell state values of the parent node are generated based on the hidden state and the cell state of its children, child sum Tree LSTM is well suited for trees with a high branching factor or whose children are unordered. Because of this phenomenon, it is a good choice for dependency trees where the number of dependents of a parent can be highly variable.

\subsection{Structured Attention}

The attention mechanism \cite{bahdanau2014neural} has been a breakthrough in neural machine translation (NMT) in recent years. This mechanism calculates how much attention the network should give to each source word to generate a specific translated word. The context vector calculated by the attention mechanism mimics the syntactic skeleton of the input sentence precisely given a sufficient number of examples. Recent work suggests that incorporating explicit syntax alleviates the burden of modeling grammatical understanding and semantic knowledge from the model \cite{grammerNMT}. However, these features are designed by evaluating the model on some downstream tasks without having any representation \cite{linzen2016assessing}.

Sentences in bio-medical texts can be comparatively quite complex. For instance, information about a protein relation sometimes extends over more than one syntactic constituent, or a modifier following a protein name sometimes names a new protein. As a consequence, most of the research uses dependency graph information as an external feature or carefully engineers more compact features extracted from the dependency tree arcs \cite{zhang2018hybrid, peng2017deep}. On the other hand, some research adopts input latent graph parsing \cite{hashimoto2017neural} as the syntax representation. 
Inducing the dependency tree in a principled manner while training allows the model to learn the internal representation of the sentence very well \cite{koo2007structured, grammerNMT}.

In our structured attention model, the input sentence is first fed to a BLSTM which gives output at each time step for each word in the sentence.
\begin{equation}
    \textbf{S} = BLSTM (x)
\end{equation}
where the term $S$ is the content annotation. Next, this $S$ is transformed into a structured annotation as a matrix through structured attention. To do this, first, we initialize three matrices $\textbf{W}_q$, $\textbf{W}_k$, $\textbf{W}_v \in \mathbb{R}^{d\times d}$ and add them as trainable model parameters. Here $d$ is the hidden dimension of the BLSTM. Then using these matrices we map $S$ into query, key and value matrices $S_q = \textbf{W}_q \textbf{S}$, $S_k = \textbf{W}_k \textbf{S}$, $S_v = \textbf{W}_v \textbf{S} \in \mathbb{R}^{n \times d}$ respectively. Here $n$ is the length of the source sentence. Next we use Kirchhoff’s matrix-tree theorem for computing marginals of non-projective dependency parsing and calculate a structured attention matrix on BLSTM output $S$ \cite{tutte1984graph}.
\begin{figure*}[ht]
\centerline{\includegraphics[width=18cm,height=10 cm]{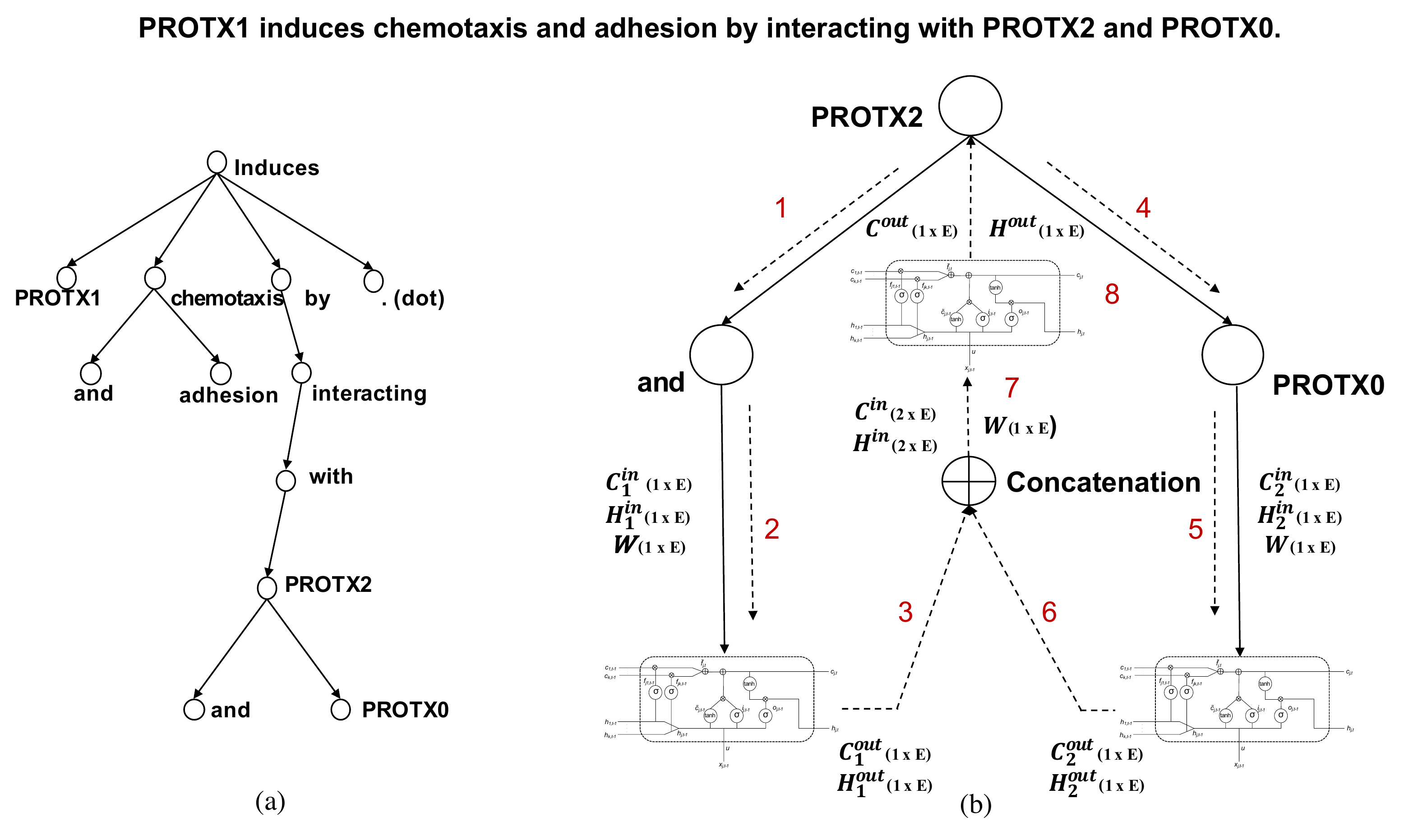}}
\caption{ \label{model}Work flow of a child sum tree LSTM on part of a dependency tree}
\label{fig}
\end{figure*}
At first, we multiply the query and key matrices to get an intermediate score matrix .
\begin{equation}
    \textbf{score}_i = S_q {S_k}^T
\end{equation}
Next, we initialize a query matrix $R_q \in \mathbb{R}^{1\times d} $ for the root node and add it as a model parameter. Following this, we multiply this $R_q$ with the key matrix $S_k$ to get a vector of length $n$
\begin{equation}
    \textbf{root} = S_k {R_q}^T
\end{equation}
Next we pick the diagonal elements of $score_i$ and add it with the $root$ vector to get a final score and then we normalize it using a partition function. Finally, we arrange this vector in the form of a block-diagonal matrix of size $n \times n$. We call this matrix $\phi$. The cell $\phi_{ij}$ means how likely the word $x_i$ is to be the parent of word $x_j$ as it captures all pairwise word dependencies. 

We are interested in selecting a soft parent word for each word and to do this we can transform the matrix $\phi$ into an attention matrix $A$ where each cell $A_{ij}$ is the posterior probability $p(\textbf{x}_i = parent(\textbf{x}_j)| \textbf{x})$. We define an adjacency matrix $z \in \{0 \times 1\}^{n \times n}$ in order to encode the source's dependency tree. We can transform our posterior into a marginal by defining it as $p(\textbf{z}_{ij} = 1 | \textbf{x}; \bm{\phi})$ which is interpreted as the probability of word $x_i$ to be the parent of word $x_j$ given the input $x$ and matrix $\phi$. So the term $A$ becomes

\begin{equation}
    A_{ij} = p(\textbf{z}_{ij} = 1 | \textbf{x}; \bm{\phi}) = \sum_{z : z_{ij}=1}^{} p(\textbf{z} | \textbf{x}; \bm{\phi})
\end{equation}
Next, we calculate the marginal of non-projective dependency structures using a framework proposed by \cite{koo2007structured} which utilizes Kirchhoff’s Matrix-Tree Theorem \cite{tutte1984graph}. In order to fill all the cells of the attention matrix $A$, we need to calculate the spanning tree from each source word in the sentence along with the probability of reaching every target node. To do this we first define a Laplacian matrix $\textbf{L} \in \mathbb{R}^{n \times n}$ as follows:

\begin{equation}
    \textbf{L}_{ij} (\bm{\phi})  = 
    \begin{cases}
      \sum\limits_{ \substack{k=1\\ k\neq j}}^{n} \exp(\bm{\phi}_{kj}), & \text{if}\ i = j \\
      -\exp(\bm{\phi}_{ij}), & \text{otherwise}
    \end{cases}
\end{equation}
Next we define another matrix $ \tilde{\textbf{L}}$ for root word selection as follows:
\begin{equation}
    \tilde{\textbf{L}}_{ij}(\bm{\phi}) = 
    \begin{cases}
      \exp(\bm{\phi}_{jj}), & \text{if}\ i = 1 \\
      \textbf{L}_{ij}(\bm{\phi}), & \text{if}\ i > 1
    \end{cases}
\end{equation}
The marginals are then calculated as,
\begin{equation}
\begin{split}
    p_1 = (1-\delta_{1j})\bigg\{ \exp({\bm{\phi}_{ij}}) \Big[ \tilde{\textbf{L}}^{-1} (\bm{\phi}) \Big]_{jj} \bigg\} \\
    p_2 = (1-\delta_{i1})\bigg\{ \exp({\bm{\phi}_{ij}})\Big[ \tilde{\textbf{L}}^{-1}
    (\bm{\phi}) \Big]_{ji} \bigg\}\\
    \end{split}
\end{equation}
\begin{equation}
    \textbf{A}_{ij} = p_1 - p_2 
\end{equation}
where $\delta_{ij}$ is the Kronecker delta. Finally the marginals for the root node is calculated as,
\begin{equation}
    \textbf{A}_{k,k} = \exp({\bm{\phi}_{k,k}})\bigg[ \tilde{\textbf{L}}^{-1} (\bm{\phi}) \bigg]_{k,1}
\end{equation}
This marginal computation is fully differentiable, thus we can train the model with the standard back-propagation algorithm \cite{goh1995back}.

\subsection{Combining the modules}
In this subsection, we combine tLSTM and structured attention as discussed above to build our final model. To the best of our knowledge no work has combined the independently produced gold standard dependency structure information with learning the structure through the model without accessing the actual dependency tree. The method described below accomplishes this fusion.

For the tree LSTM module, we first take the raw sentence and apply the Stanford dependency parser to represent it as a vector of parents where the value $j$ at index $i$ means word $x_j$ is the parent of word $x_i$ in the dependency tree. We call this vector $\textbf{P}$. Next, using this $\textbf{P}$ we compute a tree for each sentence and as an attribute we store all of its child information. This allows us to recursively traverse the entire tree if we start from the root. Apart from this, we have another matrix $\textbf{W}$ which is the embedded representation of each of the words in the sentence. Next, we pass the root of this tree and $\textbf{W}$ to a recursive module which returns a hidden state and a cell state value for the entire sentence by traversing in a tree-structured manner. Fig.~\ref{model} shows the work flow of tLSTM model on the dependency tree of a sentence. Fig.~\ref{model}(a) shows a sample dependency tree of one of the sentences from the corpus and Fig.~\ref{model}(b) shows how the hidden state and cell state of the root node of a sub-tree gets calculated. As shown in Fig.~\ref{model}(b), for a sub-tree with two children, the work flow is as follows:

While doing a traversal from the root, the tLSTM calculates the hidden state and the cell state of a node using its child hidden state, child cell state which have already been calculated recursively,

\begin{equation} \label{left}
    \textbf{H}, \textbf{C} = \textbf{tLSTM}(W, (H_i)^c,  (C_i)^c)
\end{equation}
here, $i$ represents the $i^{th}$ child, $H$ is the hidden state and $C$ is the cell state. $H$ and $C$ marked with $c$ refers to the child hidden and cell state. For our example, Eqn.~\ref{left} gets called for the word `and' and for the word `PROTX0' as leaf nodes and returns two sets of hidden state and cell state vectors. Next we concatenate the two hidden state vectors and the two cell state vectors and again apply Eqn.~\ref{left} on the resulting vector. But as a parameter, this time we pass the word vector for `PROTX1', the concatenated hidden state vector and the concatenated cell state vector. This gives us a new hidden state and cell state vectors for the word `PROTX1'. We continue to traverse the whole dependency tree in this manner, finally finishing with an encoded hidden state value $\textbf{H}_e \in \mathbb{R}^{1\times n}$ and a cell state value $\textbf{C}_e \in \mathbb{R}^{1\times n}$ for the entire tree.

For the structured attention module, we use $\textbf{W}$ as input and apply a BLSTM on it to get an output vector, $\textbf{O}$, which contains the LSTM output for each time step. Next, we pass this to the structured attention (sAttn) module which gives an attention matrix $\bm{\gamma} \in \mathbb{R}^{n \times n}$ as output.
\begin{equation}
    \bm{\gamma} = \textbf{sAttn}(\textbf{O})
\end{equation}
Next we use this $\bm{\gamma}$ with value matrix $\bm{S}_v$ to calculate the syntactic context $\textbf{C}_s \in \mathbb{R}^{n \times d}$ as follows.

\begin{equation}
    \textbf{C}_s = \gamma {S_v}
\end{equation}
Following this, we take the vectors only at the first and last index of $\gamma$ which has the entire left context as well as right context information respectively and concatenate them. We term this as $\tilde{\textbf{C}}_s$. Then we concatenate this $\tilde{\textbf{C}}_s$ with $\textbf{H}_e$ to get the final context $\textbf{M}$. 
Next, we use an MLP followed by \texttt{sigmoid} over this \textbf{M} to generate a non-linear version $\tilde{\textbf{M}}$. Finally, our model predicts a corresponding label $y$  from this $\tilde{\textbf{M}}$ as follows,
\begin{equation}
    \begin{gathered}
        p(\textbf{y} | \textbf{x}, \bm{\theta}) = \texttt{sigmoid}(\texttt{MLP}({\textbf{M}})) \\
        y_{i} = \argmax_{y} p(\textbf{y} | \textbf{x}, \bm{\theta})
    \end{gathered}
\end{equation}
\section{Experimental Analysis and Results}
In this section, we describe the results obtained with our proposed architecture. We use precision, recall and F-score as our evaluation metrics. This section also contains the detailed statistics of all of the five PPI corpora, the preprocessing steps applied to convert the problem into classification domain as well as the hyper-parameter settings of our models. In addition to that, it contains the results of the top performing models for all the corpora and extensive comparative analysis with our models. Finally, we conclude this section by giving cross corpus evaluation statistics of our architecture where we train our model on one corpus and test on another.

\begin{table}[t]
\begin{center}
\caption{\label{corpora}Basic statistics of the corpora}
\setlength\extrarowheight{3pt}
\begin{tabular}{l r r r r}
\hline
\textbf{Corpus} & \textbf{\#Positive} & \textbf{\#Negative} & \textbf{\#Sentences} \\
\hline
AIMed & $1,000$ & $4,834$ & $1,955$ \\
BioInfer & $2,534$ & $7,132$ & $1,100$ \\
IEPA & $335$ & $482$ & $486$ \\
HPRD50 & $163$ & $270$ & $145$ \\
LLL & $164$ & $166$ & $77$ \\
\hline
\end{tabular}
\end{center}
\end{table}

We evaluate our tLSTM model on five publicly available PPI corpora: AIMed \cite{aimed}, BioInfer \cite{bioinfer}, IEPA \cite{iepa}, HPRD50 \cite{hprd50} and LLL \cite{lll}. In our experiments, we use the converted version of these corpora\footnote{http://mars.cs.utu.fi/PPICorpora/} and details about these along with the conversion characteristics can be found in \cite{pyysalo2008comparative}. The statistics of the five PPI corpora are given in Table \ref{corpora}.

In order to generalize the learned model, we have modified the corpora slightly. Protein names are replaced with special symbols in each sentence, i.e., PROTX0, PROTX1 and PROTX2. Here, PROTX1 and PROTX2 are the proteins of interest and all other non-participating proteins are marked as PROTX0. For example, the following sentence ``PROTX1 induces chemotaxis and adhesion by interacting with PROTX2 and PROTX0'' indicates that PROTX1 and PROTX2 have a positive interaction. Similarly, the sentence ``PROTX0 induces chemotaxis and adhesion by interacting with PROTX1 and PROTX2'' indicates that PROTX1 and PROTX2 have a negative interaction. In this example there are three possible pairs of proteins and hence three variants of the sentence is possible. Two of them have positive interaction and one has negative interaction. In general, if a sentence has $n$ protein references, there are ${n\choose 2}$ protein pairs and hence ${n\choose 2}$ variants of the sentence.

We evaluated our model with 10-fold cross validation on each corpus allowing us to compare our results with relevant earlier works. In $k$-fold cross validation, the corpus is divided into $k$ parts, $(k-1)$ parts are training data and the other part is testing data, and is repeated $k$ times. We used StratifiedKFold from Python's Scikit-learn package which preserves the percentage of samples for each class in each fold \cite{scikit-learn}.

Table \ref{hyper} shows the detailed hyper-parameter settings used for our model. We trained our model on a GeForce GTX 1080 GPU with the `Adam' and `SGD' optimizers. All the results in the next section are reported using `SGD' as it was giving the best results. The `Learning rate decay' parameter was only used with the `SGD' optimizer. We used PyTorch 0.4 to implement our model under the Linux environment.

\begin{table}[t]
\begin{center}
\caption{ \label{hyper} The hyper-parameters used in our experiment}
\setlength\extrarowheight{3pt}
\begin{tabular}{l l}
\hline
\textbf{Hyper-parameter} & \textbf{Values} \\
\hline
Number of layers & $1 / 2$ \\
Embedding dimensions & $200$ \\
Hidden dimensions & $300/400/500$ \\
Batch size & $10/16/20$ \\
Number of epochs & $30/40/50$ \\
Dropout rate & $0.5/0.1$ \\
Learning rate & $0.001/0.015$ \\
Learning rate decay & $0.05$ \\
\hline
\end{tabular}
\end{center}
\end{table}

\begin{table*}[h]
\begin{center}
\caption{\label{results}Results (in \%) of our model (tLSTM) from 10-fold cross-validation against other methods. Bold text indicates the best performance in a column. \textbf{GK:} Graph Kernel (Airola et al., 2008). \textbf{CK:} Composite Kernel (Miwa et al., 2009). \textbf{WWSK:} Walk-weighted Subsequence Kernel (Kim et al., 2010). \textbf{NHGK:} Neighborhood Hash Graph Kernel (Zhang et al., 2011). \textbf{EDG:} Extended Dependency Graph (Peng et al., 2015). \textbf{PIPE:} Protein-protein Interaction Passage Extraction (Chang et al., 2016). \textbf{Bi-LSTM:} Bidirectional Long-Short Term Memory (Hsieh et al., 2017). \textbf{RNN + CNN:} Combination of Recurrent and Convolutional Neural Network (Zhang et al., 2018).}
\setlength\extrarowheight{3pt}
\begin{tabular}{l|c c c|c c c|c c c|c c c|c c c}
\hline
\multirow{2}{*}{\textbf{Methods}}&  \multicolumn{3}{c|}{\textit{AIMed}}& \multicolumn{3}{c|}{\textit{BioInfer}}& \multicolumn{3}{c|}{\textit{IEPA}}& \multicolumn{3}{c|}{\textit{HPRD50}}& \multicolumn{3}{c}{\textit{LLL}} \\
\cline{2-16} 
&\textbf{P} &\textbf{R} &\textbf{F1} &\textbf{P} &\textbf{R} &\textbf{F1} &\textbf{P} &\textbf{R} &\textbf{F1} &\textbf{P} &\textbf{R} &\textbf{F1} &\textbf{P} &\textbf{R} &\textbf{F1} \\

\hline
GK \cite{airola2008all}& $52.9$ & $61.8$ & $56.4$&  $56.7$ & $67.2$ &$61.3$&   $69.6$ & $82.7$ & $75.1$&   $64.3$ & $65.8$ & $63.4$&   $72.5$ & $87.2$ & $76.5$ \\
CK \cite{miwa2009protein}& $55.0$ & $68.8$ & $60.8$&  $65.7$ & $71.1$ &$68.1$&   $67.5$ & $78.6$ & $71.7$&   $68.5$ & $76.1$ & $70.9$&   $77.6$ & $86.0$ & $80.1$ \\
WWSK \cite{kim2010walk}& $61.4$ & $53.3$ & $56.6$&  $61.8$ & $54.2$ &$57.6$&   $66.7$ & $69.2$ & $67.8$&   $73.7$ & $71.8$ & $72.9$&   $76.9$ & $91.2$ & $82.4$ \\
NHGK \cite{zhang2011neighborhood}& $54.9$ & $68.5$ & $60.2$&  $59.3$ & $68.1$ &$63.4$&   $72.4$ & $79.8$ & $75.3$&   $67.8$ & $\mathbf{85.3}$ & $74.6$&   $86.2$ & $\mathbf{92.1}$ & $\mathbf{89.1}$ \\
EDG \cite{peng2015extended}& $57.3$ & $65.3$ & $61.1$&  $57.6$ & $59.9$ &$58.7$&   $69.9$ & $76.2$ & $72.9$&   $76.7$ & $83.3$ & $79.9$&   $\mathbf{92.1}$ & $78.2$ & $84.6$ \\
PIPE \cite{chang2016pipe}& $57.2$ & $64.5$ & $60.6$&  $68.6$ & $70.3$ &$69.4$&   $62.5$ & $\mathbf{83.3}$ & $71.4$&   $63.8$ & $81.2$ & $71.5$&   $73.2$ & $89.6$ & $80.6$ \\
Bi-LSTM \cite{hsieh2017identifying}& $78.8$ & $75.2$ & $76.9$&  $87.0$ & $87.4$ &$87.2$&   $-$ & $-$ & $-$&   $-$& $-$ & $-$&   $-$ & $-$ & $-$ \\
RNN + CNN \cite{zhang2018hybrid}& $52.9$ & $61.8$ & $56.4$&  $56.7$ & $67.2$ &$61.3$&   $69.6$ & $82.7$ & $75.1$&   $64.3$ & $65.8$ & $63.4$&   $72.5$ & $87.2$ & $76.5$ \\
\hline
\hline
tLSTM& $80.5$& $80.8$& $80.6$& $88.3$& $87.9$& $88.1$& $77.0$& $76.7$& $76.4$& $\mathbf{82.4}$& $82.8$& $\mathbf{82.0}$& $85.3$& $84.9$& $84.8$  \\
tLSTM + tAttn& $\mathbf{81.4}$& $\mathbf{81.9}$& $\mathbf{81.6}$& $\mathbf{88.9}$& $\mathbf{89.3}$& $\mathbf{89.1}$& $\mathbf{78.6}$& $78.7$& $\mathbf{78.5}$& $81.7$& $82.3$& $81.3$& $84.8$& $84.3$& $84.2$  \\
\hline
\end{tabular}
\end{center}
\end{table*}
Table \ref{results} shows the overall evaluation of our model in terms of precision, recall and F-score for the five PPI corpora and compares these results with the currently available state of the art models. Among these five corpora, AIMed is the most difficult as it has more noise, the sentences have nested named entities and there are many inaccurate annotations. With the AIMed corpus, we achieved a highest F-score of \textbf{81.6\%} with a significant \textbf{4.7} percentage points improvement over the previous best model. The tLSTM + tAttn model also achieved the best precision and recall scores of \textbf{81.4\%} and \textbf{81.9\%}, respectively. Our tLSTM model without attention also surpassed all of the existing models with a significant improvement in precision, recall and F-score achieving 80.5\%, 80.8\% and 80.6\%, respectively. The previous best model \cite{hsieh2017identifying} uses just the raw words and a BLSTM to capture the word context from the forward and backward directions. The second corpus that we evaluated our model on is BioInfer, the corpus with the most (9666) annotated interactions among the five. It has fewer sentences but more annotated examples than AIMed indicating that the sentences are significantly longer and contain a large number of proteins in a single sentence. With BioInfer, our tLSTM + tAttn model achieves a highest precision, recall and  F-score of \textbf{88.1\%}, \textbf{89.3\%} and \textbf{89.1\%}, respectively. Our tLSTM model without attention is the second best with precision -- 88.3\%, recall -- 87.9\% and F-score -- 88.1\%. It is to be noted that we achieved state of the art results with all evaluation metrics on these two large and complex corpora without any manual feature engineering. With the IEPA corpus, our tLSTM + tAttn model achieves a highest precision and F-score of \textbf{78.6\%} and \textbf{78.5\%}, respectively. Our model's recall score is 78.7\% which is behind only the 83.3\% of \cite{chang2016pipe} which combines PPI with a convolution tree kernel. However, their precision and F-score is low compared to both of our tLSTM and tLSTM + tAttn models. Regarding the HPRD50 corpus, our tLSTM model without attention achieves the best precision and F-score of \textbf{82.4\%} and \textbf{82.0\%} respectively. Our tLSTM + tAttn model is the second best in terms of precision and F-score. However, none of our models reached the best recall score of 85.3\% by \cite{zhang2011neighborhood} which is based on extracting the higher order substructure of the dependency graph by bit label operations on dependency graph nodes. Again, their precision and F-score is low compared to both of our models. With the LLL corpus, none of our models achieve best scores. Instead the tLSTM model without attention achieves the second best F-score of 84.8\% and the third best precision score of 85.3\%. The best recall and F-score of 92.1\% and 89.1\% is achieved by \cite{zhang2011neighborhood} which uses a neighbourhood hash graph kernel whereas \cite{peng2015extended} achieves the best precision score of 92.1\% using an extended dependency graph. An interesting aspect of our evaluation is that whenever the number of training data samples is large, no matter how complex the samples are, deep learning based methods perform very well compared to the feature based methods. With a small number of training data samples, the performance can fall short of other methods. This is what happens with LLL, the smallest corpus. Also a large number of training data samples allows the structured attention mechanism to extract the dependency information very well. That is why for comparatively large corpora, AIMed, BionInfer and IEPA, our model with attention performs best and achieves state of the art results, whereas for the two small corpora, tLSTM without attention performs better.

\begin{table}[b]
\begin{center}
\caption{\label{cross}Cross-corpus results (F-score in \%). Rows correspond to training corpora and columns to testing. Models marked with $\dagger$ represents $tLSTM$ and $\ddagger$ represents $tLSTM + tAttn$}
\setlength\extrarowheight{2.8pt}
\begin{tabular}{c c c c c c}
\hline
& \textit{AIMed} & \textit{BioInfer} & \textit{IEPA} & \textit{HPRD50} & \textit{LLL}\\
\hline
AIMed $\dagger$ & $-$ & $47.0$ & $38.6$ & $41.5$ & $34.6$ \\
AIMed $\ddagger$ & $-$ & $45.0$ & $37.9$ & $39.1$ & $33.5$ \\
BioInfer $\dagger$ & $50.8$ & $-$ & $40.8$ & $43.7$ & $35.0$ \\
BioInfer $\ddagger$ & $50.0$ & $-$ & $40.0$ & $45.5$ & $33.5$ \\
\hline
\end{tabular}
\end{center}
\end{table}

The cross corpus evaluation is inspired by the work \cite{van2008extracting} to answer the fundamental question of practical PPI extraction -- ``which corpus to be trained on in real life?''. Table \ref{cross} shows this cross corpus evaluation. Rows correspond to the training corpora and columns correspond to the test corpora. We only used AIMed and BioInfer as the training corpora and ignored the small ones because there is no point in training on small simple corpora and test on large complex corpora as suggested in \cite{peng2017deep}. It is clearly visible that the performance degrades on all of the corpora as the training and testing sets are not from the same distribution which goes against the fundamental machine learning theory about training and test sets being identically distributed. Being larger in size, the models that are trained on BioInfer perform better than the models trained on AIMed. One more interesting aspect of our evaluation is that the models without attention perform better than the models with attention. The main reason is that our structured attention captures the syntactic dependencies in the sentences and because of the two different distributions between the training and testing sets, the attention mechanism fails to capture these dependencies. Overall our cross corpus evaluation is close to the one from \cite{peng2017deep} with a slight improvement when training on BioInfer and testing on AIMed.

\section{Conclusions}
In this paper, we propose a tree recurrent neural network architecture with structured attention mechanism for the supervised PPI extraction problem. Our model gets significant improvement on two largest public PPI corpora, AIMed and BioInfer. Addition to that, our model gets state of the art result for several other small corpora too. Our experimental result shows that our tree LSTM model with structured attention is more suitable compared to traditional recurrent neural network based approaches for extracting useful features from dependency tree information of a given bio-medical text. Moreover, we believe that other linguistics features that are already proven to be useful for PPI can be included to improve the model. In future, we would like to explore the idea of leveraging other features to make our model more accurate.

\bibliographystyle{IEEEtran}
\bibliography{PPI}
\end{document}